# On Multiple User Channels with Causal State Information at the Transmitters

Styrmir Sigurjónsson and Young-Han Kim*
Information Systems Laboratory
Stanford University University
Stanford, CA 94305, USA
Email: {styrmir,yhk}@stanford.edu

*Abstract*— We extend Shannon's result on the capacity of channels with state information to multiple user channels. More specifically, we characterize the capacity (region) of degraded broadcast channels and physically degraded relay channels where the channel state information is causally available at the transmitters. We also obtain inner and outer bounds on the capacity region for multiple access channels with causal state information at the transmitters.

## I. INTRODUCTION

In his 1958 paper [1], Shannon considered a communication system where additional side information about the channel is causally available at the transmitter. (See Figure 1.) He showed

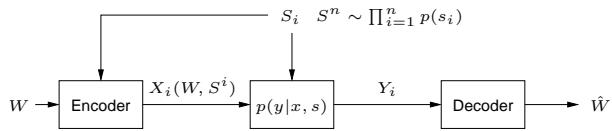

Fig. 1. Communication channel with state information causally known at the transmitter.

that the capacity of this channel can be achieved by adding a physical device in front of the channel, which depends on the current state (and the message to be sent) only. Using the modern language, we can write the capacity of this channel as

$$C = \max_{p(u)p(x|u,s)} I(U;Y)$$
$$= \max_{p(u),\, x=f(u,s)} I(U;Y), \qquad (1)$$

where $U$ is an auxilliary random variable with finite cardinality. The achievability of the rate in (1) is clear from Shannon's original argument of attaching a physical device in front of the channel. The optimality of this rate is deceptively easy to show. Indeed, by recognizing $U_i := (W, Y^{i-1}, S^{i-1})$ to be independent of $S_i$ and by observing that $X_i$ can be written as

*This work was partly supported by NSF Grant CCR-0311633.

a function of $(U_i, S_i)$, we have

$$\begin{aligned}
I(W;Y^n) &= H(Y^n) - H(Y^n|W) \\
&\leq \sum_i H(Y_i) - H(Y_i|W, Y^{i-1}) \\
&\leq \sum_i H(Y_i) - H(Y_i|W, Y^{i-1}, S^{i-1}) \\
&= \sum_i I(U_i; Y_i) \leq nC.
\end{aligned} \qquad (2)$$

It is also worth pointing out the similarity of (1) to the capacity $C'$ when the state information is *noncausally* available at the transmitter, as shown by Gel'fand and Pinsker [2] and Heegard and El Gamal [3]:

$$C' = \max_{p(u|s)p(x|u,s)} I(U;Y) - I(U;S).$$

Since the capacity $C$ for the causal case can be written as $\max_{p(u)p(x|u,s)} I(U;Y) - I(U;S)$, the loss of causality merely lies in the independence between $U$ and $S$.

In this paper, we try to extend Shannon's result to multiple user channels. Research in this direction is not new. Most notably, Steinberg [4] obtained bounds on the capacity region of degraded broadcast channels when the state information is *noncausally* available at the transmitter, although the tightness of these bounds is still open.

Causality of the state information makes the problem much easier. Recently, Steinberg [5] reported the capacity region of the degraded broadcast channel when the state information is causally available at the transmitter. In Section II, we give an independent treatment of his result by scaling the converse technique we just developed for the single-user case. In Section III, we run the same program for the physically degraded relay channel and obtain the capacity. Unfortunately, the optimality of the similar coding scheme for multiple access channels is yet to be established, and we present inner and outer bounds on the capacity region in Section IV.

## II. DEGRADED BROADCAST CHANNELS

*Definition 1:* The discrete memoryless broadcast channel with state information consists of input alphabet $\mathcal{X}$, state alphabet $\mathcal{S}$, output alphabet $\mathcal{Y}_1 \times \mathcal{Y}_2$, and a probability transition function $p(y_1, y_2|x, s)$, as in Figure 2.

We assume the state information $S^n$ is available causally to the transmitter. We also assume that the channel is physically degraded, i.e.,

$$p(y_1, y_2|x, s) = p(y_1|x, s)p(y_2|y_1).$$

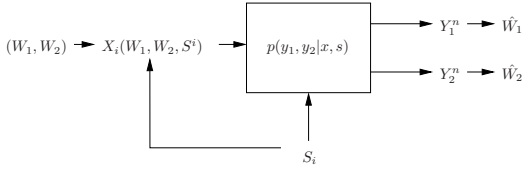

Fig. 2. Broadcast channel with state causally available at the transmitter.

*Theorem 1:* The capacity region of the degraded broadcast channel with state information available causally at the transmitter is the set of all rate pairs $(R_1, R_2)$ satisfying

$$R_1 \leq I(U_1; Y_1|U_2)$$
$$R_2 \leq I(U_2; Y_2)$$

for some joint distribution $p(u_1, u_2)p(s), x = f(u_1, u_2, s)$ where the auxiliary random variables $U_1$ and $U_2$ have finite alphabets.

*Proof:* Although the proof of the achievability can be found in [5], we repeat it here for the sake of completeness.

Fix $R_1$, $R_2$ and $p(u_2)p(u_1|u_2)p(x|u_1, u_2, s)$.

*Codebook Generation.* Generate $2^{nR_2}$ independent codewords $u_2^n(w_2)$ according to $\prod_{i=1}^n p(u_{2i})$. For each of these codewords, generate $2^{nR_1}$ independent codewords $u_1^n(w_1, w_2)$ according to $\prod_{i=1}^n p(u_{1i}|u_{2i})$. Codebook assignments are fixed and revealed to the transmitter and the receivers.

*Encoder.* To send $w_1 \in \{1, 2, \ldots, 2^{nR_1}\}$ to receiver 1 and $w_2 \in \{1, 2, \ldots, 2^{nR_2}\}$ to receiver 2, we select the corresponding codewords $u_2^n(w_2)$ and $u_1^n(w_1, w_2)$. At time $i$, upon observing the channel state $s_i$, the transmitter sends $x_i = f(u_{1,i}, u_{2,i}, s_i)$.

*Decoding.* Receiver 2 declares that a message $\hat{w}_2$ was sent if there is a unique $\hat{w}_2$ such that $u_2^n(\hat{w}_2)$ and $y_2^n$ are jointly typical; otherwise an error is declared. Receiver 1 declares that a message pair $(\hat{w}_1, \hat{\hat{w}}_2)$ was sent if there is a unique $(\hat{w}_1, \hat{\hat{w}}_2)$ such that $u_1^n(\hat{w}_1, \hat{\hat{w}}_2)$, $u_2^n(\hat{\hat{w}}_2)$ and $y_2^n$ are jointly typical; otherwise an error is declared.

Define the following events:

$$E_{2,i} = \{(U_2^n(i), Y_2^n) \in A_\epsilon^{(n)}\}$$
$$E_{1,i} = \{(U_2^n(i), Y_1^n) \in A_\epsilon^{(n)}\}$$
$$E_{1,i,j} = \{(U_2^n(i), U_1^n(j, i), Y_1^n) \in A_\epsilon^{(n)}\}$$

Without loss of generality, assume the sent messages were $w_1 = 1$ and $w_2 = 1$ and denote the probabilities that receivers 1 and 2 declare an error with $P_1^{(n)}$ and $P_2^{(n)}$, respectively. We thus have

$$P_2^{(n)} = P\left(E_{2,1}^c \bigcup \cup_{i \neq 1} E_{2,i}\right)$$
$$\leq P(E_{2,1}^c) + \sum_{i \neq 1} P(E_{2,i})$$
$$\leq \epsilon + 2^{nR_2} 2^{-n(I(U_2; Y_2) - 2\epsilon)}.$$

Hence, $P_2^{(n)} \to 0$ if $R_2 < I(U_2; Y_2) - 2\epsilon$. Similarly,

$$P_1^{(n)} = P\left(E_{1,1,1}^c \bigcup \cup_{i \neq 1} E_{1,i} \bigcup \cup_{j \neq 1} E_{1,1,j}\right)$$
$$\leq P(E_{1,1,1}^c) + \sum_{i \neq 1} P(E_{1,i}) + \sum_{j \neq 1} P(E_{1,1,j})$$

By the asymptotic equipartition property (AEP), $P(E_{1,1,1}^c) \to 0$ and $\sum_{i \neq 1} P(E_{1,i}) \leq 2^{nR_2} 2^{-n(I(U_2;Y_1) - 2\epsilon)} \to 0$ if $R_2 < I(U_2; Y_1)$. But we already have $R_2 < I(U_2; Y_2)$ and the degradedness implies $I(U_2; Y_2) < I(U_2; Y_1)$. Similarly, we can show that

$$P(E_{1,1,j}) \leq 2^{-n(I(U_1;Y_1|U_2) - 3\epsilon)}.$$

Thus $\sum_{j \neq 1} P(E_{1,1,j}) \leq 2^{nR_1} 2^{-n(I(U_1;Y_1|U_2) - 3\epsilon)} \to 0$ if $R_1 < I(U_1; Y_1|U_2) - 3\epsilon$, which completes the proof of achievability.

Let us now prove the converse. We wish to show that given any sequence of $((2^{nR_1}, 2^{nR_2}), n)$ codes $(X_i(W_1, W_2, S^i), \hat{W}_1(Y_1^n), \hat{W}_2(Y_2^n))$ such that

$$P_e^{(n)} := P(W_1 \neq \hat{W}_1 \text{ or } W_2 \neq \hat{W}_2) \to 0$$

for uniform and independent message indices $W_1$ and $W_2$, the rate pair $(R_1, R_2)$ must satisfy the conditions in the theorem. By Fano's inequality,

$$H(W_1|Y_1^n) \leq nR_1 P_e^{(n)} + 1 = n\epsilon_{1n}$$
$$H(W_2|Y_2^n) \leq nR_2 P_e^{(n)} + 1 = n\epsilon_{2n}$$

since $P_e^{(n)} \geq \max\{P_1^{(n)}, P_2^{(n)}\}$. Define the following auxiliary random variables, $U_{1i} := (W_1, Y_1^{i-1}, S^{i-1})$ and $U_{2i} := (W_2, Y_1^{i-1})$. We now have

$$nR_2 \leq H(W_2)$$
$$= I(W_2; Y_2^n) + H(W_2|Y_2^n)$$
$$\leq I(W_2; Y_2^n) + n\epsilon_{2n}$$
$$= \sum_{i=1}^n I(W_2; Y_{2i}|Y_2^{i-1}) + n\epsilon_{2n}$$
$$\leq \sum_{i=1}^n I(W_2, Y_2^{i-1}; Y_{2i}) + n\epsilon_{2n}$$
$$\leq \sum_{i=1}^n I(\underbrace{W_2, Y_1^{i-1}}_{U_{2i}}; Y_{2i}) + n\epsilon_{2n}$$
$$= \sum_{i=1}^n I(U_{2i}; Y_{2i}) + n\epsilon_{2n}.$$

Similarily,

$$nR_1 \leq H(W_1)$$

$$= I(W_1; Y_1^n) + H(W_1|Y_1^n)$$
$$\leq I(W_1; Y_1^n) + n\epsilon_{1n}$$
$$\leq I(W_1; Y_1^n|W_2) + n\epsilon_{1n}$$
$$= \sum_{i=1}^{n} I(W_1; Y_{1i}|W_2, Y_1^{i-1}) + n\epsilon_{1n}$$
$$\leq \sum_{i=1}^{n} I(\underbrace{W_1, Y_1^{i-1}, S^{i-1}}_{U_{1,i}}; Y_{1i}|\underbrace{W_2, Y_1^{i-1}}_{U_{2,i}}) + n\epsilon_{1n}$$
$$\leq \sum_{i=1}^{n} I(U_{1i}; Y_{1i}|U_{2i}) + n\epsilon_{1n}.$$

We define a random variable $Q$ independent of everything else, uniformly distributed over $\{1, 2, \ldots, n\}$ and define $U_1 = (Q, U_{1Q})$, $U_2 = (Q, U_{2Q})$, $X = X_Q$, $S = S_Q$, $Y_1 = Y_{1Q}$, and $Y_2 = Y_{2Q}$. Then, we have

$$R_1 \leq \frac{1}{n} \sum_{i=1}^{n} I(U_{1i}; Y_{1i}|U_{2i}) + \epsilon_{1n}$$
$$= \frac{1}{n} \sum_{i=1}^{n} I(U_{1i}; Y_{1i}|U_{2i}, Q=i) + \epsilon_{1n}$$
$$= I(U_{1Q}; Y_{1Q}|U_{2Q}, Q) + \epsilon_{1n}$$
$$\leq I(U_{1Q}, Q; Y_{1Q}|U_{2Q}, Q) + \epsilon_{1n}$$
$$= I(U_1; Y_1|U_2) + \epsilon_{1n},$$

and

$$R_2 \leq \frac{1}{n} \sum_{i=1}^{n} I(U_{2i}; Y_{2i}) + \epsilon_{2n}$$
$$\leq I(U_{2Q}; Y_{2Q}|Q) + \epsilon_{2n}$$
$$\leq I(U_{2Q}, Q; Y_{2Q}) + \epsilon_{2n}$$
$$= I(U_2; Y_2) + \epsilon_{2n}.$$

But it is see to see that $(U_1, U_2)$ is independent of $S$, that $X$ is a deterministic function of $(U_1, U_2, S)$, and that the joint distribution of $(X, S, Y_1, Y_2)$ is consistent with the channel $p(y_1, y_2|x, s)$. Thus, we have established the converse of the theorem. ∎

### III. PHYSICALLY DEGRADED RELAY CHANNELS

*Definition 2:* The discrete memoryless relay channel with state information consists of input alphabet $\mathcal{X}$, relay input alphabet $\mathcal{X}_1$, state alphabet $\mathcal{S}$, relay output alphabet $\mathcal{Y}_1$, output alphabet $\mathcal{Y}$, and a probability transition function $p(y, y_1|x, x_1, s)$.

In the following, we will assume that the state variable $S$ is causally available to the transmitter and the relay. We also assume that the channel is physically degraded, i.e.,

$$p(y, y_1|x, x_1, s) = p(y_1|x, x_1, s)p(y|y_1, x_1, s).$$

The main result of this section is the following:

*Theorem 2:* The capacity of the degraded relay channel with state information causally available at the transmitter and relay is given by

$$C = \max_{\substack{p(u, u_1) \\ x = f(u, s) \\ x_1 = f_1(u_1, s)}} \min\{I(U, U_1; Y), I(U; Y_1|U_1, S)\}.$$

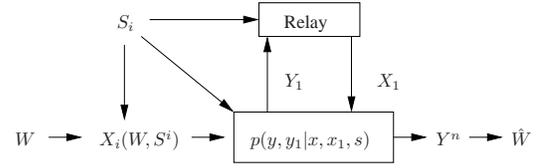

Fig. 3. Relay channel with state information available causally at the transmitter and relay

*Proof:* We first prove the achievability of the rate region. The approach follows Shannon's method of attaching physical devices [1] and the relay coding theorem by Cover and El Gamal [6], which transforms the original relay channel into one with auxilliary inputs $U$ and $U_1$.

More specifically, we use the block Markov encoding. A sequence of $B - 1$ messages, $w_i \in \mathcal{W}$, each selected independly and uniformly over $\mathcal{W}$, are to be sent over the channel in $nB$ transmissions. Within each block of length $n$, the sender and the relay use a doubly-indexed set of codewords

$$\mathcal{C} = \{(u^n(w|t), u_1^n(t)) : w \in \{1, \ldots, 2^{nR}\},$$
$$t \in \{\emptyset, 1, \ldots, 2^{nR_0}\}\}$$

Fix a probability distribution $p(u, u_1)$, $x = f(u, s)$, $x_1 = f_1(u_1, s)$.

*Codebook Generation:* Generate at random $2^{nR_0}$ independent $n$-sequences $u_1^n(t)$, $t \in \{1, \ldots, 2^{nR_0}\}$, each drawn according to $\prod_{i=1}^{n} p(u_{1i})$. For each $u_1^n(t)$ sequence, generate $2^{nR}$ conditionally independent $u^n(w|t)$ sequences drawn according to $\prod_{i=1}^{n} p(u_i|u_{1i})$. This defines the random codebook $\mathcal{C}$.

For each message $w \in \{1, \ldots, 2^{nR}\}$ assign an index $t(w)$ at random from $\{1, \ldots, 2^{nR}\}$. The set of messages with the same index form a bin $\mathcal{T}_t \subset \mathcal{W}$. The codebook and bin assignments are revealed to all parties.

*Encoding:* Let $w(b) \in \{1, \ldots, 2^{nR}\}$ be the new index to be sent in block $b$, and assume that $w(b-1) \in \mathcal{T}_{t(b)}$. The encoder then selects $u^n(w(b)|t(b))$. At time $i$ in block $b$, upon receiving $s_i(b)$, the sender sends $x_i(b) = f(u_i(w(b)|t(b)), s_i(b))$. The relay will have an estimate $\hat{w}(b-1)$ of the previous index $w(b-1)$. Assume that $\hat{w}(b-1) \in \mathcal{T}_{\hat{t}(b)}$. Then upon receiving $s(b)$ the relay encoder sends $x_{1i} = f_1(u_{1i}(\hat{t}(b)), s_i(b))$.

*Decoding:* We assume that at the end of block $b-1$, the receiver knows $(w(1), \ldots, w(b-2))$ and $(t(1), \ldots, t(b-1))$ and the relay knows $(w(1), \ldots, w(b-1))$ and consequently $(t(1), \ldots, t(b))$. The decoding procedures at the end of block $b$ are as follows:

1) With $(t(b), y_1^n(b), s^n(b))$, the relay estimates the message of the transmitter as $\hat{w}(b)$ if there exists a unique $\hat{w}(b)$ such that $(u^n(\hat{w}(b)|t(b)), u_1^n(t(b)), y_1^n(b), s^n(b))$ are jointly typical.

It can be shown that $\hat{w}(b) = w(b)$ with arbitrarily small probability of error, if

$$R < I(U; Y_1 | U_1, S)$$

and $n$ sufficiently large.

2) The receiver declares that $\hat{t}(b)$ was sent if there exists exactly one $\hat{t}(b)$ such that $(u_1^n(\hat{t}(b)), y^n(b))$ is jointly typical. It can be shown that $\hat{t}(b) = t(b)$ with arbitrarily small probability of error, if

$$R_0 < I(U_1; Y) \qquad (3)$$

and $n$ sufficiently large.

3) Assuming that $t(b)$ is decoded succesfully at the receiver, then $\hat{w}(b-1)$ is declared to be the index sent in block $b-1$ if there is a unique $\hat{w}(b-1) \in \mathcal{T}_{t(b)}$ that is jointly typical with $y^n(b-1)$. It can be shown that if $n$ is sufficiently large and if

$$R < I(U; Y | U_1) + R_0 \qquad (4)$$

then $\hat{w}(b-1) = w(b-1)x$ with arbitrarily small probability of error. Combining (3) and (4) yields the condition $R < I(U, U_1; Y)$.

Let us now prove the converse for any rate-$R$ code with encoding functions $X_i(W, S^i)$ and $X_{1i}(S^i, Y^{i-1})$. We define the auxiliary random variables $U_i = (W, Y^{i-1}, S^{i-1})$ and $U_{1i} = (Y_1^{i-1}, S^{i-1})$. It is easy to see that $(U_i, U_{1i})$ is independent of $S_i$ and that $X_i$ and $X_{1i}$ are deterministic functions of $(U_i, S_i)$ and $(U_{1i}, S_i)$, respectively. We have

$$I(W; Y^n) = \sum_{i=1}^n I(W; Y_i | Y^{i-1})$$
$$= \sum_{i=1}^n H(Y_i | Y^{i-1}) - H(Y_i | Y^{i-1}, W)$$
$$\leq \sum_{i=1}^n H(Y_i) - H(Y_i | Y^{i-1}, Y_1^{i-1}, W, S^{i-1})$$
$$= \sum_{i=1}^n I(U_i, U_{1i}; Y_i)$$

$$I(W; Y^n) \leq I(W; Y^n, Y_1^n | S^n)$$
$$= \sum_{i=1}^n H(Y_i, Y_{1i} | Y^{i-1}, Y_1^{i-1}, S^n)$$
$$\quad - H(Y_i, Y_{1i} | Y^{i-1}, Y_1^{i-1}, S^n, W)$$
$$= \sum_{i=1}^n H(Y_i, Y_{1i} | Y^{i-1}, Y_1^{i-1}, S^n)$$
$$\quad - H(Y_i, Y_{1i} | Y^{i-1}, Y_1^{i-1}, S^n, W, X_i, X_{1i})$$
$$\leq \sum_{i=1}^n H(Y_i, Y_{1i} | Y_1^{i-1}, S^{i-1}, S_i)$$
$$\quad - H(Y_i, Y_{1i} | X_i, X_{1i}, S_i)$$
$$\leq \sum_{i=1}^n H(Y_i, Y_{1i} | Y_1^{i-1}, S^{i-1}, S_i)$$

$$\quad - H(Y_i, Y_{1i} | X_i, X_{1i}, S_i, Y^{i-1}, Y_1^{i-1}, S^{i-1}, W)$$
$$\leq \sum_{i=1}^n H(Y_i, Y_{1i} | U_{1i}, S_i) - H(Y_i, Y_{1i} | U_i, U_{1i}, S_i)$$
$$= \sum_{i=1}^n I(U_i; Y_i, Y_{1i} | U_{1i}, S_i).$$

Fano's inequality and the use of the usual time-sharing random variable will show

$$R \leq \min \{ I(U, U_1; Y), I(U; Y, Y_1 | U_1, S) \} + \epsilon_n.$$

Combining the following Markov relationship

$$U \to (S, U_1, Y_1) \to Y$$

with the degradedness of the channel and the fact that $X$ and $X_1$ are functions of $(U, S)$ and $(U_1, S)$, we can easily show that $I(U; Y, Y_1 | U_1, S) = I(U; Y_1 | U_1, S)$. This completes the proof of the converse. ∎

## IV. MULTIPLE ACCESS CHANNELS

*Definition 3:* The discrete memoryless multiple access channel with state information consists of input alphabet $\mathcal{X}_1, \mathcal{X}_2$, state alphabet $\mathcal{S}$, output alphabet $\mathcal{Y}$, and a probability transition function $p(y | x_1, x_2, s)$; see Figure 4.

Again we consider the case when the state variable $S$ is causally available to the transmitters.

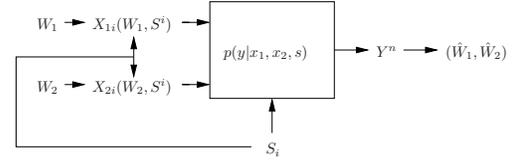

Fig. 4. Multiple access channel with state causally available to the transmitters.

Define the region $R_{p(u_1, u_2)}^{\text{mac}}$ to be the convex hull of all rate pairs $(R_1, R_2)$ satisfying

$$R_1 \leq I(U_1; Y | U_2)$$
$$R_2 \leq I(U_2; Y | U_1)$$
$$R_1 + R_2 \leq I(U_1, U_2; Y)$$

over all $p(u_1, u_2)$, $x_1 = f_1(u_1, s)$, $x_2 = f(u_2, s)$. Similarly, define $R_{p(u_1)p(u_2)}^{\text{mac}}$ to be the convex hull of all rate pairs $(R_1, R_2)$ satisfying the same set of inequalities over all $p(u_1, u_2)$, $x_1 = f_1(u_1, s)$, and $x_2 = f(u_2, s)$.

For the multiple access channel case, the capacity theorem is yet to be established, mostly because of the coupling of two auxiliary random variables in the proof of the converse. Here we give bounds on the capacity region instead.

*Theorem 3:* Let $C^{\text{mac}}$ denote the capacity region of the multiple access channel with state causally available at both transmitters. Then $R_{p(u_1)p(u_2)}^{\text{mac}} \subset C^{\text{mac}} \subset R_{p(u_1, u_2)}^{\text{mac}}$.

*Proof:* Following the same arguments as before, it is not hard to show that $C^{\text{mac}} \subset R_{p(u_1, u_2)}^{\text{mac}}$; hence, we skip the details.

For the lower bound, we first fix $p(u_1)p(u_2)$, $f_1(u_1, s)$, and $f_2(u_2, s)$.

*Codebook Generation*. Generate $2^{nR_1}$ independent codewords $u_1^n(w_1)$, $w_1 \in \{1, 2, \ldots, 2^{nR_1}\}$, generating each element i.i.d. $\sim \prod_{i=1}^n p(u_{1i})$ and $2^{nR_2}$ independent codewords $u_2^n(w_2)$, $w_2 \in \{1, 2, \ldots, 2^{nR_2}\}$, generating each element i.i.d. $\sim \prod_{i=1}^n p(u_{2i})$. These codewords form the codebook, which is revealed to the senders and the receiver.

*Encoding*. To send message indices $w_1 \in \{1, 2, \ldots, 2^{nR_1}\}$ and $w_2 \in \{1, 2, \ldots, 2^{nR_2}\}$, select the corresponding codewords $u_1^n(w_1)$ and $u_2^n(w_2)$. At time i, upon receiving $s_i$, transmitter 1 sends $x_{1,i} = f_1(u_{1i}(w_1), s_i)$ and transmitter 2 acts similarly.

*Decoding*. Let $A_\epsilon^{(n)}$ denote the set of typical $(U_1^n, U_2^n, Y^n)$ sequences. If there exists a unique pair $\hat{W}_1$ and $\hat{W}_2$ such that $(U_1^n(\hat{W}_1), U_2^n(\hat{W}_2), Y^n) \in A_\epsilon^{(n)}$ declare that $(\hat{W}_1, \hat{W}_2)$ was sent, otherwise declare an error.

We now analyze the probability of error. Define the events,

$$E_{i,j} = \{(U_1^n(i), U_2^n(j), Y^n) \in A_\epsilon^{(n)}\}.$$

Without loss of generality assume the messages sent were $W_1 = 1$ and $W_2 = 2$, the probability of error becomes,

$$P_e^{(n)} = P\left\{E_{1,1}^c \bigcup \cup_{(i,j) \neq (1,1)} E_{i,j}\right\}$$
$$\leq P(E_{1,1}^c) + \sum_{i \neq 1} P(E_{i,1}) + \sum_{j \neq 1} P(E_{1,j})$$
$$+ \sum_{(i,j) \neq (1,1)} P(E_{i,j})$$

From the AEP, $P(E_{1,1}^c) \to 0$. For the second term we get

$$P(E_{i,1}) = P\{(U_1^n(i), U_2^n(1), Y^n) \in A_\epsilon^{(n)}\}$$
$$= \sum_{(u_1^n, u_2^n, y^n) \in A_\epsilon^{(n)}} p(u_1^n) p(u_2^n, y^n)$$
$$\leq \sum_{(u_1^n, u_2^n, y^n) \in A_\epsilon^{(n)}} 2^{-n(H(U_1)-\epsilon)} 2^{-n(H(U_2,Y)-\epsilon)}$$
$$\leq 2^{n(H(U_1,U_2,Y)+\epsilon)} 2^{-n(H(U_1)-\epsilon)} 2^{-n(H(U_2,Y)-\epsilon)}$$
$$= 2^{-n(H(U_1)+H(U_2,Y)-H(U_1,U_2,Y)-3\epsilon)}$$
$$= 2^{-n(I(U_1;U_2,Y)-3\epsilon)}$$
$$= 2^{-n(I(U_1;Y|U_2)-3\epsilon)}$$

where the last equality follows since $U_1$ and $U_2$ are independent. Similarily for $j \neq 1$,

$$P(E_{1,j}) \leq 2^{-n(I(U_2;Y|U_1)-3\epsilon)},$$

and for $(i,j) \neq (1,1)$

$$P(E_{i,j}) \leq 2^{-n(I(U_1,U_2;Y)-4\epsilon)}.$$

We can now write the probability of error as,

$$P_e^{(n)} \leq P(E_{1,1}^c) + 2^{nR_1} 2^{-n(I(U_1;Y|U_2)-3\epsilon)}$$
$$+ 2^{nR_2} 2^{-n(I(U_2;Y|U_1)-3\epsilon)}$$
$$+ 2^{n(R_1+R_2)} 2^{-n(I(U_1,U_2;Y)-4\epsilon)}$$

which tends to zero if the conditions of the theorem are met. ∎

## V. Concluding Remarks

We characterized the capacity region of a few simple multiple user channels with state information, in single-letter formulas. The value of this result may lie in expanding the set of toy examples in network information theory and thus giving a glimpse on the structure of the theory. Questions still remain on other multiple user channels and on the cost of causality of state information.

## References

[1] C. E. Shannon, "Channels with side information at the transmitter," *IBM Journal of Research and Development,* vol. 2, pp. 289–293, 1958.
[2] S. I. Gel'fand and M. S. Pinsker, "Coding for channel with random parameters," *Problems of Control and Information Theory,* vol. 9, no. 1 pp. 19–31, 1980.
[3] C. Heegard and A. A. El Gamal, "On the capacity of computer memories with defects," *IEEE Trans. Info. Theory,* vol. 29, pp. 731–739, September 1983.
[4] Y. Steinberg, "On the broadcast channel with random parameters," in *Proc. IEEE ISIT 2002 (Lausanne, Switzerland),* p. 225, June 2002.
[5] Y. Steinberg, "Coding for the degraded broadcast channel with random parameters, with causal and non-causal side information," accepted for publication in *IEEE Trans. Info. Theory.*
[6] T. M. Cover and A. A. El Gamal, "Capacity theorems for the relay channel," *IEEE Trans. Info. Theory,* vol. IT-25, pp. 572-582, Sep. 1979.